\documentclass[twocolumn,prl,amsmath,amssymb]{revtex4}
\usepackage{graphicx}
\usepackage{color}

\def\t{\theta}

\def\s{\sigma}

\def\sin{\rm sin}

\def\ul{\underline}

\def\be{\begin{equation}}
\def\ee{\end{equation}}
\def\ba{\begin{eqnarray}}
\def\ea{\end{eqnarray}}

\def\br{\begin{array}}
\def\er{\end{array}}
\def\bc{\begin{center}}
\def\ec{\end{center}}
 
\def\gsim{\raisebox{-.5ex}{\rlap{$\sim$}} \raisebox{.5ex}{$>$}}

\def \GeV{~{\rm GeV}}
\def \TeV{~{\rm TeV}}
\begin{document}
\setcounter{footnote}{0}
\renewcommand{\thefootnote}{\fnsymbol{footnote}}
 
\title{The Impact of Non-zero $\theta_{13}$ on \\ Neutrino Mass and Leptogenesis in a SUSY SO(10) Model} 

\author{Swarup Kumar Majee}
\email{swarup.majee@gmail.com}

\affiliation{ Department of Physics, National Taiwan University, Roosevelt Road, Taipei, Taiwan 10617}

\begin{abstract} 
\noindent

The recent measurement of the reactor angle as $\sin^2 2\theta_{13} =
0.092 \pm 0.016(stat) \pm 0.005(syst)$ come from the Daya Bay
collaboration. Evidence of nonzero $\theta_{13}$ was also there at T2K,
MINOS and Double Chooz experiments. We study the implication of these recent data on neutrino mass matrix and consequently on leptogenesis in a supersymmetric SO(10) model. To explain the smallness of neutrino mass, in general, we require a heavy Majorana neutrino which is a natural candidate in SO(10) model. In minimal SO(10) model, the symmetry breaking scale or the right-handed neutrino mass scale is close to the GUT scale. It is not only beyond the reach of any present or future collider search but the lepton asymmetry generated from its decay is in conflict with the gravitino constraint as well as unable to fit the neutrino data. We show that addition of an extra fermion singlet can accommodate the observed recent neutrino data in a supersymmetric SO(10) model. This model can generate the desired lepton asymmetry and provide TeV scale doubly-charged Higgs scalars to be detected at LHC. 
\end{abstract}

\maketitle
 
\renewcommand{\thefootnote}{\fnsymbol{footnote}} 
\normalsize 
 
\renewcommand{\thesection}{\Roman{section}} 
\setcounter{footnote}{0} 
\renewcommand{\thefootnote}{\arabic{footnote}} 

The standard model (SM) of particle physics is based on the gauge 
group $SU(3)_C\times SU(2)_L\times U(1)_Y$, where C, L
and Y respectively stand for colour, left-handed and hypercharge quantum numbers. 
In the SM, due to conservation of lepton family number and absence
of any right-handed (RH) counterpart, left-handed neutrinos are
massless. However, it is well established that neutrino flavour
oscillates which require neutrinos to be massive. One can generate tiny 
neutrino mass either by various type of seesaw mechanisms
\cite{seesaw} or via loop corrections \cite{zee} going beyond the standard model. Neutrino oscillations 
can be parametrized using two mass squared differences, three
mixing angles and one CP-phase of Pontecorvo-Maki-Nakagawa-Sakata matrix \cite{pmns}.
The third mixing angle, namely the reactor angle, was little known
till date.  

In neutrino physics, a breakthrough measurement of the third mixing
angle $\t_{13}$ come from Daya Bay experiment \cite{dayabay}, and is 
confirmed by RENO experiment \cite{reno}.  A more than $5\s$
measurement given by Daya Bay as $\sin^2 2\t_{13} = 0.092 \pm
0.016(stat) \pm 0.005(syst)$ and the corresponding value from the
RENO experiment is $\sin^2 2\t_{13} = 0.113 \pm 0.013(stat)
\pm 0.019(syst)$. Their central values are very close to what predicted
in ref. \cite{fogli, valle} using the combined data set of T2K \cite{t2k} and
MINOS \cite{minos}, earlier in the mid of last year, with more than
$3\sigma$ evidence. There was also a similar result from Double
Chooz experiment \cite{dchooz} as well. All these new data, thus, make
the neutrino mass matrix much more constrained and their consequences
to other areas of physics. This new measurement led to prediction and
implication of the $\theta_{13}$ angle in different ways \cite{q13}.

The main ingredient of  see-saw mechanism is heavy Majorana 
neutrino, which is a natural candidate in the left-right symmetric 
$\bf {SO(10)}$ grand unified theories (GUTs) \cite{lr}. Embedding supersymmetry (SUSY)
in such a model have some good features \emph{like} protecting the
Higgs mass from radiative correction that appear due to the huge
difference between the weak and unification scales and help to
have a good unification of gauge couplings at the GUT scale. We, thus, work here 
in a scenario of left-right symmetric (LR) supersymmetric $\bf {SO(10)}$
GUT model. In order to keep the gauge coupling unification intact at the
GUT scale, any intermediate scale, here the LR-symmetry breaking scale,
has to be very close to the GUT-scale. However, such a heavy symmetry
breaking scale or an equally heavy Majorana neutrino is beyond the
reach of any present or future collider analysis as well as is unable to
produce the low energy light neutrino data. 

Observed baryon asymmetry of the universe is another interesting
problem. A popular explanation is to generate baryon
asymmetry via sphaleron process from lepton asymmetry \cite{yana, ruba}. The later 
generally can be produced through the C and CP-violating
out-of-equilibrium decay of heavy Majorana neutrinos, which is a
member of ${\bf SO(10)}$-GUT model and also responsible to explain the
tiny neutrino mass.

In the standard thermal leptogenesis, with heavy hierarchical
right-handed neutrino spectrum, the CP-asymmetry and the mass of the
lightest right-handed Majorana neutrino are correlated. In order to have the
correct order of light neutrino mass-squared differences, there is a 
lower bound on the mass of the right-handed neutrino, $M_N \gsim
~10^9\GeV$ \cite{di}, which implies a reheating temperature $\gsim 10^9\GeV$. 
This will lead to an excessive gravitino production and conflicts with
the observed data as discussed below.

Gravitino, being the lightest and stable, is a suitable dark matter
candiate in a R-parity conserving SUSY. In the post-inflation era, 
these gravitino are produced in a thermal bath due to annihilation or
scattering processes of different standard particles. The relic
abundance of gravitino is proportional to the reheating temperature of
the thermal bath. One can have the right order of relic dark matter
abundance only if the reheating temperature is bounded to below $10^7
\GeV$ \cite{lepto, khlopov}.

In this article we work in left-right symmetric SUSY $\bf {SO(10)}$ GUT
model, rich with an extra $\bf {SO(10)}$ singlet lepton per
generation \cite{e6, vm, barr}. This extra singlet lepton is a natural
member in $E_6$ and many other models. The issue is earlier addressed
 in different context \cite{mpr, pr}. However, here, we are to 
 accommodated the recent neutrino data in this model. In addition, we
 discuss the impact of these new data on 
other related phenomenology.  Our analysis, in a single model, is
able to explain various issues \emph{like} the light neutrino masses
and their mixing as measured in recent experiments, have an exact
unification of different gauge couplings at the GUT scale, have a 
low intermediate scale or a lighter right-handed Majorana neutrino 
as well as to generate right amount of lepton asymmetry to explain the
observed baryon asymmetry of the universe without being in conflict
with the gravitino constraint. 

The spontaneous symmetry breaking prescription of ${\bf SO(10)}$ group in
our model is as follows -- 
At the GUT scale ${\bf SO(10)}$  is spontaneously broken with a
simultaneous vacuum expectation value (vev) to the ${\bf \Phi^{54}}$ of along the direction
singlet under the Pati-Salam group (${\cal G}_{PS}$) ${\bf SU(2)_L \times
SU(2)_R\times SU(4)_C}$ \cite{ps}  and to the singlet direction
under the left-right gauge group (${\cal G}_{LR}$) ${\bf SU(2)_L
\times SU(2)_R \times U(1)_{(B-L)}\times  SU(3)_C}$ in the ${\cal G}_{PS}$ multiplet $(1, 1, 15)$
contained in a ${\bf \Phi_{(1)}^{210}}$ of  ${\bf SO(10)}$. At
this stage D-parity remains intact and both the gauge couplings of
${\bf SU(2)_L}$ and ${\bf SU(2)_R}$ are equal, $g_L=g_R$ \cite{dpar}.  At the next
step a vev to the D-Parity odd singlet, also contained in
${\bf \Phi_{(2)}^{210}}$ of ${\bf SO(10)}$, breaks the D-parity.  To break the
LR-symmetry at the next step we assign vev to the RH doublets 
${\bf \chi_R \oplus \overline {\chi}_R} \subset {\bf {16}_H \oplus
  \overline {16}_H}$, however the subtlety of this breaking will be
discussed later in the sections while at the last step electroweak
symmetry is broken by a ${\bf {10}_H}$-plet.

The effective Lagrangian at the intermediate, LR-symmetry
breaking, scale is 
\begin{eqnarray} 
{\cal L}_Y =  & Y \overline {\psi}^{\bf{16}}_L \psi^{\bf{16}}_R\Phi^{\bf{10}} 
+ f{\psi^{\bf{16}}_R}^{\rm T}\tau_2\psi^{\bf{16}}_R \overline{\Delta^{\bf{126}}_R} \nonumber \\  
& + F\overline{\psi^{\bf{16}}_R}T^{\bf{1}}\chi^{\bf{16}}_R + \mu {T^{\bf{1}}}^{\rm T} T^{\bf{1}}+H.c. .
\label{yuklm}
\end{eqnarray} 
The interacting superpotential to the scalar fields at the
intermediate scale is given by

\begin{eqnarray} 
W =  & M_{\Delta_R}\Delta^{\bf{126}}_R\overline {\Delta^{\bf{126}}_R} + M_{\chi_R}\chi^{\bf{16}}_R\overline {\chi^{\bf{16}}_R} \nonumber \\  
& +\lambda~\overline{\Delta^{\bf{126}}_R}\chi^{\bf{16}}_R\chi^{\bf{16}}_R +  \lambda^{*}~ \Delta^{\bf{126}}_R
\overline{\chi^{\bf{16}}_R}\overline{\chi^{\bf{16}}_R} .
\label{supot}
\end{eqnarray} 

where $\psi_{L/R}$ are left-/right-handed lepton doublets, while the
superscript ${\bf 16}$ stands to represent that they belong to the
${\bf 16}$-plet of ${\bf SO(10)}$ representation and so on,  and 
$T$, not to be confused with the superscript ${\rm T}$  for the transposed
field, the fermion singlet field, one for each generation. 
The introduction of the scalar field ${\bf \Delta \subset 126}$, we
can justify from the scalar interaction terms as follows.

We have assigned a vev to the right-handed doublet component only 
${ \langle {\chi_R^{\bf{16}}}^0 \rangle =  \langle {\overline
    {\chi_R^{\bf{16}}}^0} \rangle = v_{\chi}}$. 
However, the vanishing F-term conditions give us,  
\be
\langle {\Delta_R^{\bf{126}}}^0 \rangle = v_R   = - \lambda ~ \frac{v_{\chi}^2}{M_{\Delta_R}}. \;\;
\label{vr}
\ee 

We, thus, have a large induced vev to the neutral component of the
triplet scalar ${\Delta_R^0}$ or ${\overline {\Delta}^0_R}$, once the
neutral doublet component ${\chi^0_R}$ gets a vev.  For example, with
a lighter RH-triplet mass $M_{\Delta} \simeq 100\GeV - 1\TeV$,  it is
possible to have $v_R \simeq 10^{9} - 10^{12}\GeV$ for $v_{\chi} =
10^6- 10^7\GeV$, assuming $\lambda \sim {\cal O}(1)$.  Since, $v_R \gg v_{\chi}$, the spontaneous
symmetry breaking of the group $SU(2)_R\times U(1)_{B-L} \to U(1)_Y$
takes place at a higher scale generating large RH Majorana neutrino
masses $M_N \gg M_X$. This will lead to a small $N_i - T_j$ mixings,
which is a crucial point needed to establish the out-of equilibrium
conditions for leptogenesis.

In this model the neutral fermions per generation are a left-handed
neutrino $\nu$, a right-handed neutrino $N$, both of
which are member of ${\bf 16}$-plet of ${\bf SO(10)}$,  and a 
sterile neutrino, $T$.  From the Yukawa interaction, Eq.(\ref{yuklm}),
we see in the $(\nu, N, T)$ basis the $3\times 3$ mass matrix is given by 
\begin{eqnarray}
M_{\nu} =
\left(\begin{array}{ccc} \nu & N^c & T \end{array}\right)_L 
\left(\begin{array}{ccc} 
 0 & m_D & 0 \\  
m_D^{\rm T} & M_N & M_X \\
0 & M_X^{\rm T} & \mu   
\end{array}\right)
\left(\begin{array}{c} \nu^c  \\   N \\  T \end{array}\right)_L.~
\label{matrix}
\end{eqnarray} 

Here the $N-T$ mixing matrix arises through the {\em vev} of the
RH-doublet field with 
\be
M_X = Fv_{\chi},~~{\rm where}~~ v_{\chi}=\langle \chi^0_R\rangle, 
\label{mx}
\ee
and the RH-Majorana neutrino
mass is generated by the induced {\em vev} of the RH-triplet
with 
\be
M_N = f v_R,
\label{mn}
\ee
where, $v_R$ is given in eq.(\ref{vr}) as described above. The {\em vev} of the weak
bi-doublet $\Phi(2,2,0,1) \subset 10_H$ of ${\bf SO(10)}$ yields
the Dirac mass matrix for neutrinos, 
\be
m_D = Y \langle\Phi^0 \rangle.
\label{md}
\ee

In our model different mass scales hierarchy is $M_N \gg M_X \gg \mu
\gg m_D$. With this hierarchical mass spectrum \cite{barr}, the eigenvalues of the 
mass matrix in eq. (\ref{matrix}) is given by 
\begin{eqnarray}
m_{\nu} & \sim & - {m_D}\,\left[M_X^{-1}~\mu~
(M_X^{\rm T})^{-1}\right]\,m_D^{\rm T}, 
\label{double1}
\end{eqnarray}
\begin{eqnarray}
M_T & \sim &\mu - \frac{M_X^2}{M_N}.
\label{double2}
\end{eqnarray}

Here, we see that the light neutrino masses satisfy a double see-saw
structure. It may be noted that the mass
matrix structure in eq. (\ref{matrix}) ensures that the type-I
see-saw contribution is absent  and $M_N$ remains
unconstrained by the light neutrino masses. This freedom in $M_N$
-- a hallmark of the model -- is vital to ensure adequate
leptogenesis.

In order to satisfy both the neutrino data as well as to generate the
required amount of lepton asymmetry in this ${\bf SO(10)}$ model, we 
note that the mass matrix $\mu$ can be obtained using eq.(\ref{double1})
once we know the mass matrices $m_\nu$, $m_D$ and $m_X$. Our strategy 
is described as follows:

To construct the Dirac mass matrix $m_D$, here we work in a basis in
which the down-quark and charged lepton mass matrices are diagonal. 
The entire mixings in the quark and lepton sectors, thus, can be
ascribed to the mass matrices of the up-type quarks and the neutrinos,
respectively. On the otherhand, in $\bf {SO(10)}$ model, the quark-lepton symmetry
\cite{ps} relates the neutrino Dirac mass matrix $m_D$ to its counterpart
in the up-quark sector. We, therefore, obtain $m_D$ using the quark
masses and the Cabibbo-Kobayashi-Maskawa mixing angles, upto $\cal
O$(1) effects due to RG evolution. Using the data enlisted
in the Particle Data Group \cite{pdg} for the CKM
matrix elements, its Dirac phase, and the running masses of the
three up-type quarks, namely, $m_u$ = 2.5 MeV,  $m_c$ = 1.29 GeV,
$m_t$ = 172.9 GeV, we have,
\be
m_D\simeq M_U = V_{CKM}^{\dagger}{\rm ~diag}(m_u, m_c, m_t) V_{CKM}, 
\label{dirac} 
\ee
where we have used the CKM phase  $\delta_{CKM} = 1.2$ 
radian, and the quark mixing angles $\sin\theta^q_{12} =
0.2253$, $\sin\theta^q_{23} = 0.041$, and $\sin\theta^q_{13} =
0.0041$. The Dirac neutrino mass matrix is fixed by the underlying
quark-lepton symmetry of $SO(10)$. Neglecting small RG
corrections, it is taken to be approximately equal to the
up-quark mass matrix.

Next, we see that the matrix $M_X$ is determined through 
eq.(\ref{mx}). However, the $3\times 3$ coupling matrix $F$ is
completely arbitrary. To minimize the number of independent
parameters, we take the matrix F to be real and diagonal. Here, we
choose, for example, $M_X \equiv ~{\rm diag} (0.15, 0.5,
0.8) \times v_\chi$ with $v_\chi = 10^6 \GeV$. Once a vev to the RH
doublet $\chi_R^{\bf 16}$ is chosen, we can have a definite induced
vev to the RH triplet $\Delta_R^{\bf 126}$ using eq.(\ref{vr}). Not to
mention, due to same reason like $M_X$, we choose $f \sim {\rm diag}(0.1, 0.5,
0.9)$ to obtain, via eq.(\ref{mn}), the $M_N$ mass matrix of ${\cal O}(10^{10})\GeV$. 
This will, for a general $M_X$, lead to a $N_i-T_j$ mixing given by 
\be
\sin\theta_{ij} = \frac{M_{X_{ij}}}{M_{N_i}}\,.
\ee
{\hspace*{-3mm}}\begin{table}[t]
 \begin{tabular}{|c|c|c|c|c|c|}\hline
 &$\delta m^2$ &$\Delta m^2$&${\rm sin}^2\theta_{12}$&${\rm sin}^2\theta_{23}$&${\rm
   sin}^2\theta_{13}$\\
 &$/10^{-5}{\rm eV}^2$ &$/10^{-3}{\rm eV}^2$&&&$/ 10^{-2}$\\
 \hline
 bfv&7.58 & 2.35 &0.312 &0.42&2.5\\\hline
 1$\sigma$&7.32 - 7.80& 2.26 - 2.47&0.296 - 0.329&0.39 - 0.50&1.8 - 3.2\\\hline
 2$\sigma$&7.16 - 7.99 &2.17 - 2.57 &0.280 - 0.347&0.36 - 0.60&1.2 - 4.1
   \\ \hline
   3$\sigma$& 6.99 - 8.18& 2.06 - 2.67&0.265 - 0.364& 0.34 - 0.64&0.5
   - 5.0\\ \hline
 \end{tabular}
\caption{ Ranges for mixing parameters obtained in Ref.\cite{fogli}}.
 \label{fit}
\end{table}

In neutrino physics, breakthrough measurement of the
third mixing angle $\t_{13}$ come from different experiments. Its 
evidence come in the mid last year from T2K and MINOS. Their combined
data predicted non-zero $\theta_{13}$ \cite{fogli,valle} with more
than $3\sigma$ evidence. Recently, Daya Bay as well
as RENO experiment come up with a more than $5\sigma$
measurement. However, their central value is very close to what
predicted in \cite{fogli,valle}. To construct the neutrino mass
matrix we use the combined neutrino mixing data, including the 
recent T2K and MINOS results, given in Table-\ref{fit} from
Ref.\cite{fogli} with the new reactor flux estimate.  In addition,
here, we assume all other CP-phases in the lepton sector, except the
one through the CKM matrix in the quark sector,  to be zero.
Assuming the lightest neutrino mass eigenvalue ($m_1$ for normal
hierarchy and $m_3$ for the inverted hierarchy) to be zero, we
obtained two other mass eigenvalues using different values of $\delta
m^2 (=m_2^2 -m_1^2)$ and $\Delta m^2 (= m_3^2 - (m_2^2 + m_1^2)/2)$
from Table-\ref{fit}. We construct $m_{\nu}$ from these mass eigenvalues via
the PMNS matrix, $U_{PMNS}$ as 
\be
m_{\nu} = U_{PMNS}^{\rm T}  {\rm ~diag}(m_1, m_2, m_3)U_{PMNS}\;\;.
\label{numass}
\ee

\begin{figure}[b]
  \centering
{\hspace*{-3mm}}  \includegraphics[width=0.50\textwidth,height=0.22\textheight,angle=0]
   {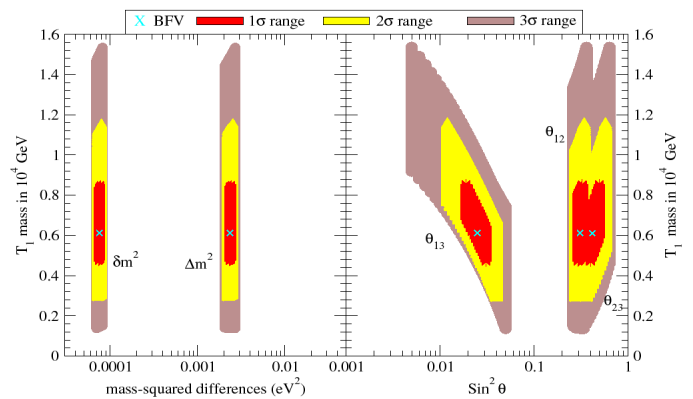}
\caption{{\ul {Normal Hierarchy}}: \em  $T_1$ mass corresponding to the best fit
  value (cyan-cross) of neutrino data and its variation for
  $1\sigma$,  $2\sigma$  and $3\sigma$ are shown by the red (central
  dark) , yellow (whitish) and brown (outer dark) areas.}
  \label{f:mass-NH}
\end{figure}
\begin{figure}[t]
  \centering
{\hspace*{-3mm}}  \includegraphics[width=0.50\textwidth,height=0.22\textheight,angle=0]
   {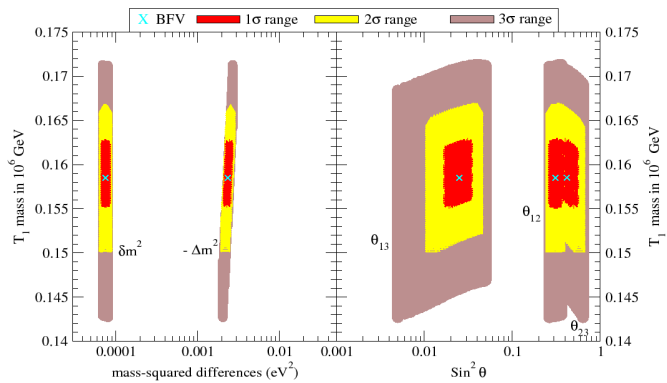}
\caption{{\ul {Inverted Hierarchy}}: \em $T_1$ mass corresponding to the best fit
  value (cyan-cross) of neutrino data and its variation for
  $1\sigma$,  $2\sigma$  and $3\sigma$ are shown by the red (central
  dark) , yellow (whitish) and brown (outer dark) areas.}
  \label{f:mass-IH}
\end{figure}

With the knowledge of $m_{\nu}$, $m_D$ and $M_X$ we then use
the inverse see-saw mass formula, from eq.(\ref{double1}), to
obtain elements of the matrix $\mu$ for both normal and inverted
hierarchical light neutrino masses. The $\mu$, $M_X$ and
$M_N$ matrices are used in eq. (\ref{double2}) to compute the mass
eigenvalues of the singlet fermions and their mixings. Thus the two input
matrices $M_N$ and $M_X$ (chosen diagonal) -- eq.  (\ref{matrix}) --
completely determine the singlet neutrino, $T_i$, masses and
mixings consistent with the recent data on the light neutrino mass
spectrum and their mixing. Out of three eigenvalues only one,
denoting it as $T_1$, is above the threshold energy to decay into
$l\phi$. We have shown the variation of the allowed $T_1$ mass in
Fig. \ref{f:mass-NH} for a normal hierarchical light neutrino mass and
in Fig. \ref{f:mass-IH} for the case of inverted
hierarchy. For normal hierarchy, best fit values given in
Table -\ref{fit} corresponds to a $T_1$ mass equals to $0.6124 \times
10^4 \GeV$ and is denoted by the cross (cyan) in
Fig. \ref{f:mass-NH}.  Here, we see that $T_1$ mass varies 
between $(0.4709 - 0.8526)\times 10^4\GeV$ (brown-central dark) for
$1\sigma$, $0.2874\times 10^4 - 0.1152\times 10^5 \GeV$ (yellow -
whitish) for $2\sigma$ and $0.1388\times 10^4 - 0.1533\times 10^5
\GeV$ (red - outer dark) for $3\sigma$ allowed experimental neutrino data.
A similar analysis is shown for the inverted hierarchy is given
in Fig. \ref{f:mass-IH}. With the same $M_X$ and $M_N$, for inverted
mass hierarchy light neutrino case, the best fit values of Table
-\ref{fit} corresponds to a $T_1$ mass equals to $0.1585 \times
10^6 \GeV$ and is denoted by the cross (cyan) in Fig. \ref{f:mass-IH}.
In this case, $T_1$ mass varies between $(0.1556 - 0.1624)\times
10^6\GeV$ (brown-central dark) for $1\sigma$, $(0.1504 - 0.1662)\times
10^6\GeV$ (yellow -whitish) for $2\sigma$ and $(0.1427 - 0.1713)\times
10^6\GeV$  (red - outer dark) for $3\sigma$ allowed experimental
neutrino data.

We now discuss to check if or not the predicted mass spectrum for $T_1$,
obtained using the neutrino data, is able to generate require amount
of lepton asymmetry. The mass scale $T_1$ is well-below the condition
on reheating temperature come from the gravitino overproduction. This
model, thus, is in good agreement with the current limit on the dark
matter relic abundance. 

\begin{figure}[tbh]
  \centering
  \includegraphics[width=0.45\textwidth,height=0.10\textheight,angle=0]
   {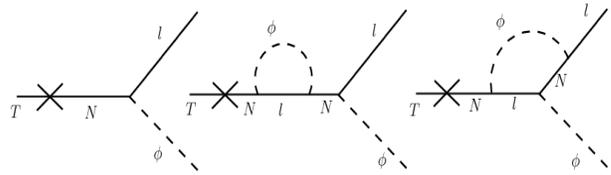}
\caption{\em The tree and one-loop  contributions to the decay of $T_1$ that generates the lepton asymmetry.}
  \label{f:feyndiag}
\end{figure}

The singlet fermions  decay through their mixing, controlled by the ratio
$M_X/M_N$,  with the $N_i$. The latter, which have masses ${\cal
O}(10^{10})$ GeV and are off-shell, decay to a final $l \phi$
state, where $l$ is a lepton doublet and $\phi$ the standard
Higgs boson. This two-step process -- for which a typical tree diagram is
depicted in Fig. \ref{f:feyndiag} -- results in a lepton asymmetry of the
correct order. Because of the large value of $M_N >> M_X$, a small
$T_i-N_i$ mixing is naturally permitted which in turn guarantees
out-of- equilibrium condition to be realised near temperature $T\simeq
M_T$.
%
\begin{figure}[b]
  \centering
\includegraphics[width=0.22\textwidth,height=0.18\textheight,angle=-90]
   {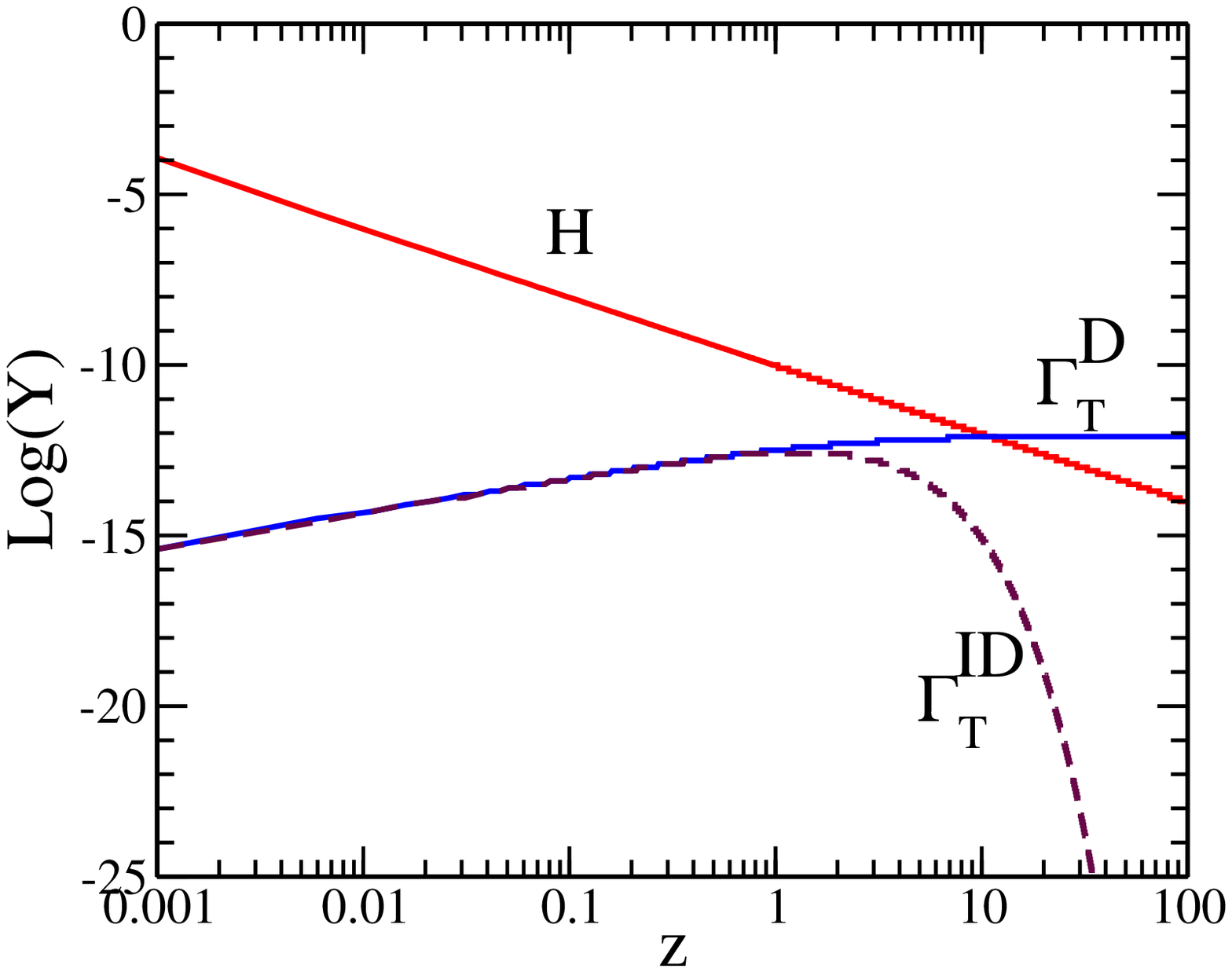}
\includegraphics[width=0.22\textwidth,height=0.18\textheight,angle=-90]
   {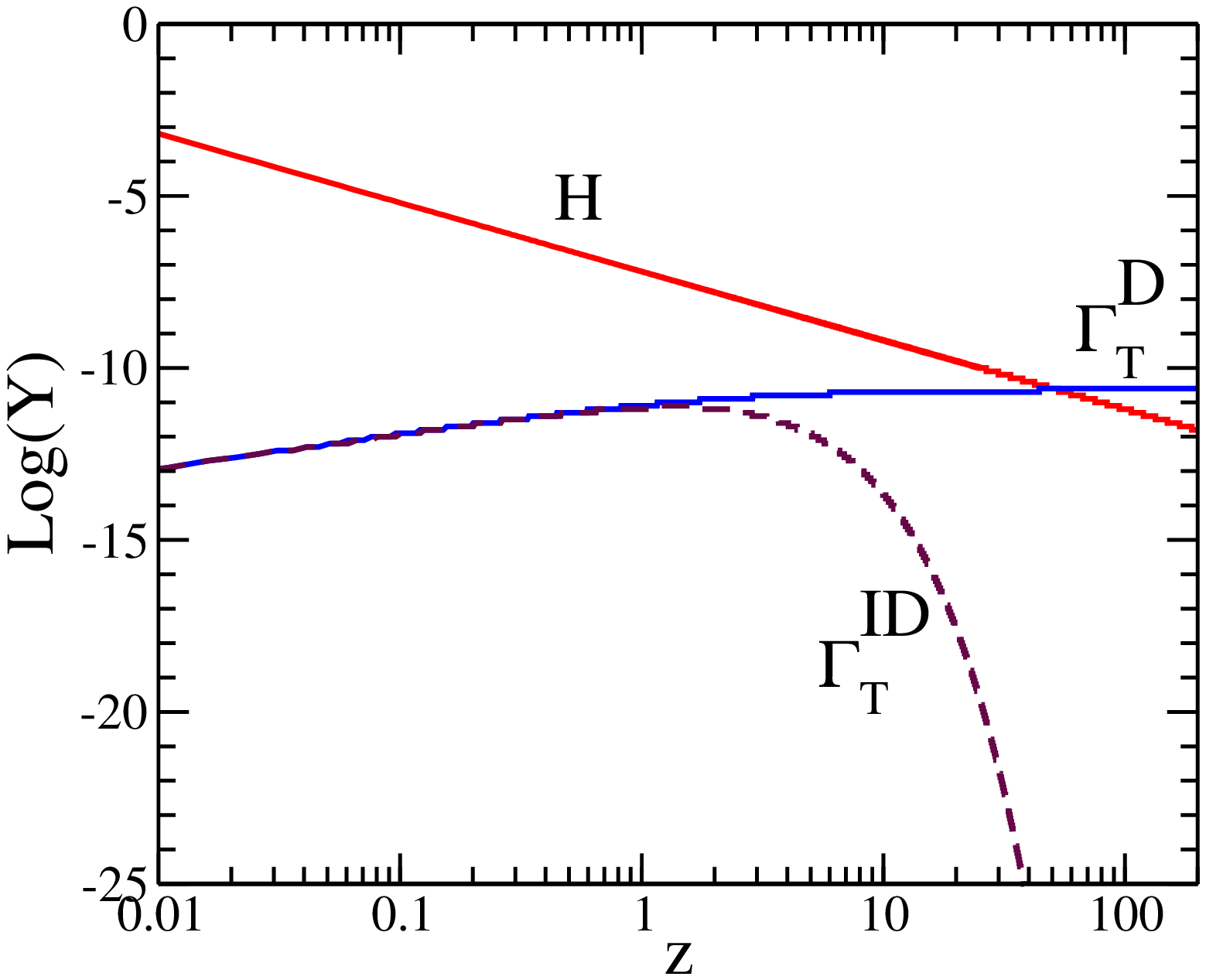}
\caption{\em 
The decay and inverse-decay rate of $T$ are compared with the Hubble
expansion rate, $H$, as a function of $z$, for the best fit values of
neutrino data only are shown for the {\ul {Normal Hierarchy}} (Left) and
\ul {Inverted Hierarchy} (Right). }
\label{f:decayvsz}
\end{figure}

Below we discuss various solutions of the Boltzmann equations. These 
determine the number densities in a co-moving volume $Y_{T} = n_{T}/n_S$ and $Y_L =
n_L/n_S$, where $n_L$ and $n_S$ are respectively the number
densities of leptons and the entropy. We can read the equations as -
\begin{eqnarray}
\frac{dY_{T}}{dz}&=&-\left(Y_{T}-Y_T^{eq}\right) \left[ \frac{\Gamma_D^T}
{zH(z)}+\frac{\Gamma_s^T}{zH(z)} \right],\nonumber \\
\frac{dY_{L}}{dz}&=&\epsilon_T \frac{\Gamma_D^T}{zH(z)} \left( Y_{T}-Y_T^{eq}
\right)-\frac{\Gamma_W^\ell}{zH(z)}Y_L.
\label{BE}
\end{eqnarray}
where $\Gamma_D^T$, $\Gamma_s^T$ and $\Gamma_W^\ell$ represent
the decay, scattering, and wash out rates, respectively, that
take part in establishing a net lepton asymmetry. We refrain from
presenting their detailed expressions here and due to negligible
contribution from supersymmetric processes \cite{nosusy}, we have not
included them. The Hubble expansion rate $H(z)$, where $z = M_{T}/T$,
and the CP-violation parameter are given by
\begin{eqnarray}
H(z)&=&\frac{H(M_T)}{z^2}, \,\,\,\, H(M_T)=1.67 g_*^{1/2}
\frac{M_T^2}{M_{pl}},\nonumber \\
\epsilon_T &=& \frac{\Gamma (T\rightarrow l
\phi) - \Gamma (T\rightarrow \bar{l}
\phi^*)}{\Gamma (T\rightarrow l
\phi) + \Gamma (T\rightarrow \bar{l}
\phi^*)}.
\label{hubble_eps}
\end{eqnarray}

\begin{figure}[t]
{\hspace*{-8mm}}  \includegraphics[width=0.40\textwidth,height=0.40\textheight,angle=-90]
   {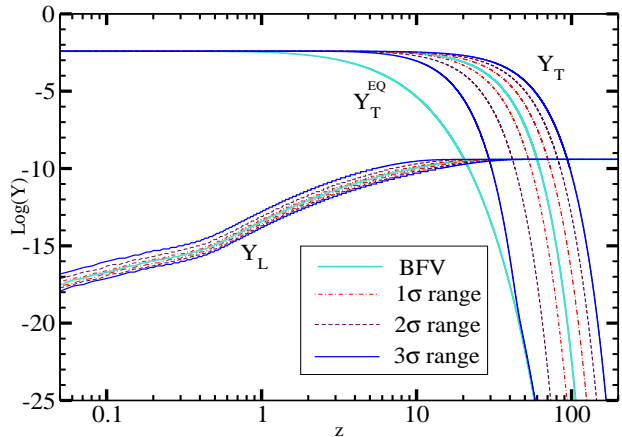}
\caption{
\ul{Normal Hierarchy}: \em The comoving density of $T$ -- $Y_T$ --  and the leptonic asymmetry  
-- $Y_L$ -- as a function of $z$ is shown. Corresponding to the best
fit neutrino data, $Y_L$, $Y_{T}$ and $Y_{T}^{EQ}$ are shown in
the central (cyan) lines and the dot-dashed (red), dashed (maroon) and solid (blue)
boundary lines corresponds to $1\sigma$, $2\sigma$ and $3\sigma$ ranges.}
\label{f:leptasy_nh}
\end{figure}

\begin{figure}[t]
{\hspace*{-8mm}}  \includegraphics[width=0.40\textwidth,height=0.40\textheight,angle=-90]
   {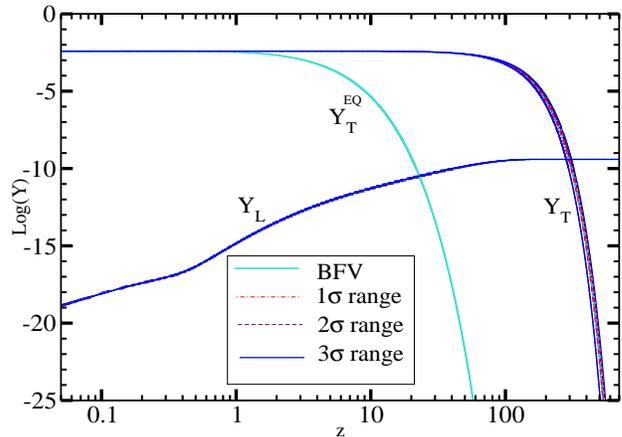}
\caption{\ul{Inverted Hierarchy}: \em Same as in {\rm Fig. \ref{f:leptasy_nh}}}
\label{f:leptasy_ih}
\end{figure}

In Fig.-\ref{f:decayvsz}, we have shown the variation of the decay and
inverse decay rate of $T_1$ for both normal (left) and inverted
(right) hierarchical light neutrino cases. Corresponding to the best
fit value mass spectrum, the figure clearly shows how the
out of equilibrium condition are satisfied to generate the lepton
asymmetry. We have shown the lepton asymmetry production results in
Fig.-\ref{f:leptasy_nh} and in Fig.-\ref{f:leptasy_ih}. We assume that
in the very initial stages the number densities $T_i, \,i=1,2,3$, are
zero. $T_1$ decay through the channel $l\phi$ to produce the lepton asymmetry.  One important point
to note here is that in this process of leptogenesis, reheating
temperature is consistent with the gravitino
constraint. In the figure, corresponding to the best fit values of neutrino data,
$Y_L$, $Y_{T_1}$ and $Y_{T_1}^{EQ}$ are plotted as the central (cyan)
lines. The effect of $1\sigma$, $2\sigma$ and $3\sigma$ ranges of
neutrino data on the evolution of $Y_L$ and $Y_{T_1}$ are shown
respectively with the dot-dashed (red), dashed (maroon) and solid (blue)
boundary lines.  It is seen from the figure that although it is
perturbed at a lower value of $z$ but as the universe expands 
$Y_L$ achieves the right order ($\sim 10^{-10}$) starting off
from a vanishing initial value while that for $Y_{T_1}$ are well
separated. However, for the inverted hierarchy case, effect 
for $1\sigma$, $2\sigma$ and $3\sigma$ are overlapping due to a 
relatively monochromatic and large $T_1$ mass.
%
Now, we just comment on how to achieve a unification of 
gauge couplings in this $\bf {SO(10)}$ model. As mentioned earlier,
Higgs multiplets {\bf 210} and {\bf 54} are utilised to
break the symmetry at $M_U$. Within the {\bf 210} there are two
components which develop vevs; one breaks $\bf {SO(10)}$ to
${\cal G}_{3221}$ while the other is responsible for D-parity
breaking. The $\bf {SU(2)_R}\times U(1)_{B-L}$ symmetry is broken by
the induced vev $\sim 10^{11} \GeV$, which is also responsible for the 
masses of the $N_i$, of RH triplets in ${\bf 126} \oplus {\bf \overline{\bf 126}}$. 
The last step of breaking carried out by the  
weak bi-doublet  in {\bf 10}. With the analysis of the gauge couplings
RG evolution we determine the intermediate mass scales. An
intermediate scale at $M_R \sim 10^{9-11}$ GeV can be obtained through
the introduction of effective $\rm {dim}.5$ operators scaled by the
Planck mass, $M_{Pl}$ \cite{d5}. It is interesting to note that both
{\bf 210} and {\bf 54} are necessary for viable SUSY ${\bf SO(10)}$ breaking
pattern and consequently the resulting two $\rm {dim}.5$ operators
appear to alleviate the problem of leptogenesis under gravitino
constraint:  
\ba
{\cal L}_{NRO} = -{1\over 2M_U}
Tr\left[F_{\mu\nu}(\eta_1\Phi_{210}+ \eta_2 \Phi_{54})F^{\mu\nu}\right].
\label{dim5}
\ea
The reason behind this is that above interaction lead to finite
corrections to the gauge couplings at the GUT-scale so that the
gauge couplings of left-right gauge group emerge from
one effective GUT-gauge coupling.  The
upshot of this is that with these additional contributions 
it is possible to lower $M_R$ to as low as
$10^9 \to 10^{11}$ GeV as required in this model. The grand
unification scale is as large as  $M_U \sim
10^{17-18}$ GeV and the model predicts a stable proton for all
practical purposes. Another way to achieve this gauge coupling
unification is to introduce some additional scalar multiplet at the
intermediate scale \cite{pr}.

Finally, we comment on the experimental prospect of doubly charged
scalar of this model at LHC or  ILC \cite{pr}.  Using the D-parity
mechanism in this model we make the RH-triplets in ${\bf
{126}_H\oplus {\overline {126}}_H}$ carrying $B-L= \pm 2 $.
Their masses are from $100 \GeV$ to a few
TeV. This light triplet scalar comes out as a necessary condition to enhance  
the induced {\em vev}, $v_R$ or to break the LR gauge
symmetry at high scale. Consequently, we have heavy RH Majorana
neutrinos as well as the corresponding gauge bosons. This forbids
$\Delta_R^{\pm\pm}$ to decay into right-handed gauge bosons. However,
after being produced via Drell-Yann process at LHC, these
doubly-charged scalars will decay to fermions to be detected at LHC. 

In conclusion, in view of the recent neutrino data we 
have presented a left-right symmetric SUSY $\bf {SO(10)}$ model. This
model is capable to solve a multi-dimensional problems.  This model
has the following features
\begin{itemize}

\item By virtue of its construction, it is consistent with the most
  recent neutrino masses and mixing angles obtained at MINOS, T2K,
  Daya Bay, RENO experiments.

\item  We have discussed the variation of the $T_1$ mass due to
  $1\sigma$, $2\sigma$, $3\sigma$ variation for both the normal and
  inverted hierarchy light neutrino cases. 

\item It generates a correct lepton asymmetry via the decays of $\bf
  {SO(10)}$ singlet neutrino with a mass scale to be consistent with
  the gravitino constraint. 

\item It can also have a good unification of different gauge
couplings at the GUT scale.

\item The model is also a source of light doubly-charged scalar. It's
  mass range is within the reach of the LHC. 
\end{itemize}

{\bf{Acknowledgements}} 

We are thankful to Mina K. Parida and Amitava
Raychaudhuri for valuable comments. The work is
partially supported by NSC 100-2811-M- 002-089.

\end{document}